\documentclass[aps,prl,twocolumn,amsmath,amssymb]{revtex4}
\usepackage{amsfonts}
\usepackage{graphicx}
\usepackage{amsbsy}
\usepackage{hyperref}
\usepackage{dcolumn}
\usepackage{bm}

\begin{document}

\textbf{Comment on ``Scaling in Plateau-Plateau Transition: A
Direct Connection of Quantum Hall Systems with Anderson
Localization Model''} \vspace{0.5cm}

In a remarkable experiment on quantum criticality in the quantum
Hall regime at finite temperatures ($T$) Wanli Li {\em et.al.}
\cite{Wanli09} recently reinvestigated the scaling law
\begin{equation}
 \left( \frac{\partial R_{xy}}{\partial B} \right)_{max} =
 \left(\frac{T}{T_s}\right)^{-\kappa}
\end{equation}
which was originally predicted by the renormalization theory of
the quantum Hall effect more than two decades
ago.\cite{Pruisken88} Here, $T_s$ is an arbitrary $T$ scale
determined by the microscopics of the sample. The universal
critical exponent $\kappa$ equals the ratio $p/2\nu$ with $\nu$
denoting the localization length exponent and $p$ the inelastic
scattering exponent. An experimental estimate $\kappa = 0.42$ was
obtained which is the same value as previously found by H.P.Wei
{\em et.al.}\cite{Wei88} Unlike the experiments by H.P.Wei {\em
et.al.}, however, Wanli Li {\em et.al.} also investigated the
finite size scaling behavior
\begin{equation}
 \left( \frac{\partial R_{xy}}{\partial B} \right)_{max} =
 \left(\frac{L}{L_s}\right)^{\frac{1}{\nu}}
\end{equation}
with $L$ denoting the physical sample size and $L_s$ an arbitrary
microscopic length. This latter behavior is expected to dominate
the experiment when $T$ is low enough such that the inelastic
scattering length $L_\phi \propto T^{-p/2}$ exceeds the value of
$L$.\cite{Pruisken88} Eqs (1) and (2) have been used to extract
the individual values $\nu=2.4$ and $p=2$ which according to Wanli
Li {\em et.al.} are the ``commonly accepted" exponent values of
the quantum Hall plateau transitions. Here, the phrase ``commonly
accepted" refers to the same exponent value $\nu=2.4$ that
previously has been extracted from numerical experiments conducted
on the disordered free electron gas. This phrase, however, is
incorrect otherwise.

Eqs (1) and (2) clearly involve the effects of the
electron-electron interaction which has been the most challenging
objective in this field of research for many years. In an
extensive series of theoretical investigations by the authors
\cite{PruiskenBurmistrov07} it was shown that the infinitely
ranged Coulomb interaction between the electrons, present in
realistic samples, completely invalidate any approach to the
laboratory experiment based on Fermi liquid type of ideas. A
unifying scaling diagram for the interacting electron gas was
obtained indicating that quantum criticality in the quantum Hall
regime generally falls into two different universality classes
with distinctly different symmetries and exponent values dependent
on the range of the electron-electron interaction. The first
class, termed the Fermi liquid universality class, is associated
with finite ranged interactions. The predicted exponent values are
$\nu=2.4$ and $p=1.35 \pm 0.15$ or $\kappa=0.29 \pm 0.04$ which is
clearly at odds with the experiment. The second class, termed the
$\mathcal F$-invariant universality class, is associated with
infinitely ranged electron-electron interactions such as the
Coulomb potential. Even though the numerical exponent values of
the latter are presently unknown, both from the theoretical side
and the numerical simulations point of view, one nevertheless has
rigorous bounds on the exponent $p$ given by $1 < p < 2$. Given
the aforementioned best experimental value of $\kappa =0.42$ one
therefore concludes that the localization length exponent of the
true electron gas is in the range $1.2 < \nu < 2.4$ which is
entirely non-Fermi liquid like.

In view of these as well as numerous other advances made in the
theory of localization and interaction effects the individual
values for $\nu$ and $p$ reported by Wanli Li {\em et.al} appear
to be highly controversial. In particular, pursuing finite size
scaling experiments based on Eq. (2) means in practice that one
compares the data taken from a number of different samples with a
varying size $L$. However, along with varying $L$ one also finds
that the absolute length scale $L_s$ and temperature scale $T_s$
for scaling are sample dependent and vary from sample to sample in
uncontrolled manner. This sample dependence, clearly displayed in
the experimental data of Wanli Li {\em et.al.}, fundamentally
complicates any attempt to disentangle the critical exponents
$\nu$ and $p$ based on Eq. (2).

Despite this flaw in the analysis of experimental data Wanli Li
{\em et.al.} nevertheless reported experimental estimates that
numerically agree with the Fermi liquid value $\nu=2.3$ known from
numerical simulations. Rather than pursuing numerical agreement
with ``commonly accepted" exponent values, however, it is more
important to keep an open mind especially where outstanding
critical phenomena are concerned that are not well understood
theoretically and inaccessible on the computer.

One of us (A.M.M.P) is indebted to D.C. Tsui for sharing the
experimental data prior to publication.

\vspace{0.2cm} \noindent A.\,M.\,M.\,Pruisken$^1$ and
I.\,S.\,Burmistrov$^{2}$\\
{\small ${}^1$Institute for Theoretical Physics, University
of Amsterdam, \\
\hphantom{${}^1$}Valckenierstraat 65, 1018 XE Amsterdam, The
Netherlands\\
${}^2$L.\,D.\,Landau Institute for
Theoretical Physics,\\
\hphantom{${}^2$}Kosygina street 2, 117940 Moscow, Russia }

\end{document}